\documentclass[twocolumn,aps,prb
groupedaddress
]{revtex4-2}
\usepackage[colorlinks,bookmarks=false,citecolor=blue,linkcolor=red,urlcolor=blue]{hyperref}
\usepackage{epsfig}
\usepackage{tikz}
\usepackage{graphicx}
\usepackage{physics}
\usepackage{dcolumn}
\usepackage{bm}
\usepackage{ mathrsfs }

\DeclareMathOperator{\arcsinh}{arcsinh}
\usepackage{amsmath,amssymb}
\DeclareRobustCommand{\stirling}{\genfrac\{\}{0pt}{}}
\makeatletter
\renewcommand*\env@matrix[1][*\c@MaxMatrixCols c]{%
  \hskip -\arraycolsep
  \let\@ifnextchar\new@ifnextchar
  \array{#1}}
\makeatother

\usepackage{appendix}

\begin{document}

\title{The operator growth hypothesis in open quantum systems}

\author{N.~S.~Srivatsa}
\affiliation{Department of Physics, King's College London, Strand WC2R 2LS, UK}

\author{Curt von Keyserlingk}
\affiliation{Department of Physics, King's College London, Strand WC2R 2LS, UK}


\begin{abstract}
The operator growth hypothesis (OGH) is a technical conjecture about the behaviour of operators -- specifically, the asymptotic growth of their \emph{Lanczos coefficients} -- under repeated action by a Liouvillian. It is expected to hold for a sufficiently generic closed many-body system. When it holds, it yields bounds on the high frequency behavior of local correlation functions and measures of chaos (like OTOCs). It also gives a route to numerically estimating response functions. Here we investigate the generalisation of OGH  to open quantum systems, where the Liouvillian is replaced by a Lindbladian. For a quantum system with local Hermitian jump operators, we show that the OGH is modified: we define a generalisation of the Lanczos coefficient and show that it initially grows linearly as in the original OGH, but experiences exponentially growing oscillations on scales determined by the dissipation strength. We see this behavior manifested in a semi-analytically solvable model (large-$q$ SYK with dissipation), numerically for an ergodic spin chain, and in a solvable toy model for operator growth in the presence of dissipation (which resembles a non-Hermitian single-particle hopping process). Finally, we show that the modified  OGH connects to a fundamental difference between Lindblad and closed systems: at high frequencies, the spectral functions of the former decay algebraically, while in the latter they decay exponentially. This is an experimentally testable statement, which also places limitations on the applicability of Lindbladians to systems in contact with equilibrium environments. 
\end{abstract}

\maketitle

\begin{figure}[t]
\includegraphics[width=\linewidth]{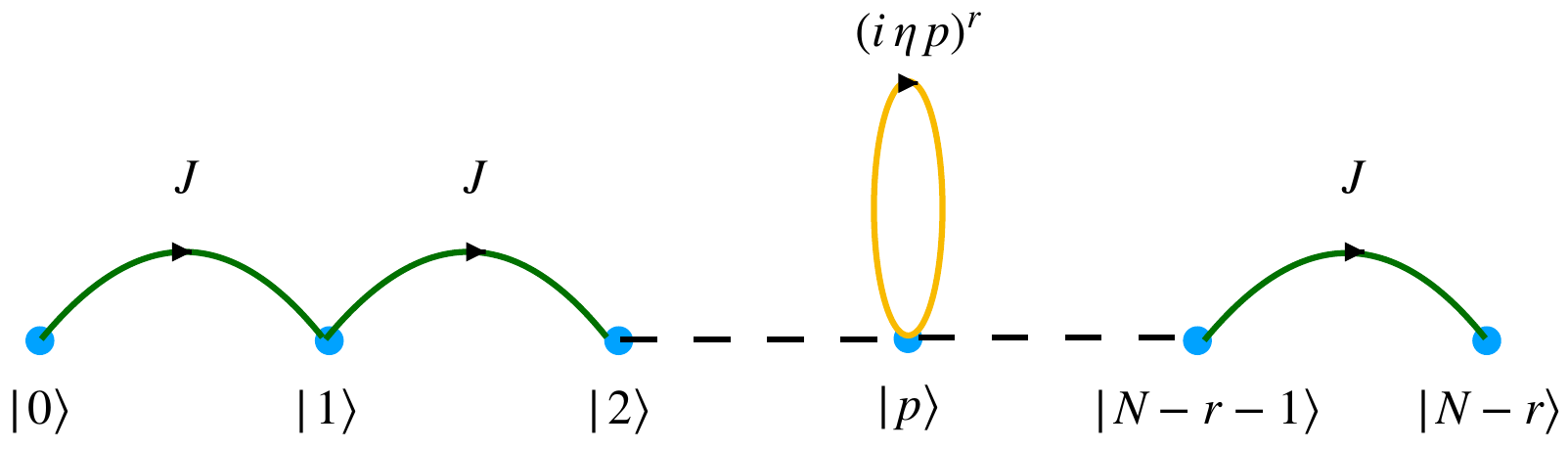}
\centering
\caption{A simplified toy model showing one possible trajectory for the growth of an initial operator $|0\rangle$ under $N$ repeated applications of the effective Lindbladian $\mathscr{L_\textrm{eff}}=\mathscr{L}_{\textrm{u}}+\mathscr{L}_{\textrm{d}}$, where the action of the unitary part $\mathscr{L}_{\textrm{u}}$ is to hop to the right with amplitude $J$ and the action of the dissipative part $\mathscr{L}_{\textrm{d}}$ is to stay on any site $|p\rangle$ with amplitude $i\,\eta\,p$  . The example trajectory shows hopping to the right for $p$ steps , dissipating $r$ steps at site $|p\rangle$ and hopping the remaining $N-p-r$ steps to land at site $|N-r\rangle$.}\label{toy1}

\end{figure}
\section{Introduction}
What are the universal signatures of thermalising or ergodic quantum systems? A number of such signatures have been proposed. The eigenstate thermalisation hypothesis conjectures a universal form for the matrix elements of local observables in the energy eigenbasis of an ergodic closed many-body Hamiltonian \cite{eth1,eth2}. Likewise the `dip-ramp-plateau' structure of the so-called spectral form factor is believed to be generic for ensembles of ergodic systems \cite{Bertini,Amos}. More recent work indicates that ergodicity is also encoded in operator dynamics and universal statements about the decay properties of out-of-time-ordered correlation (OTOC) functions \cite{roberts,jonay,khemani1,khemani2,curt1,curt2}. When ergodic systems have conserved densities, these tend to obey the laws of hydrodynamics and there is an interesting interplay between this observation and those above on operator dynamics \cite{khemani2,curt2} which has inspired new numerical algorithms for approximating many-body dynamics \cite{daoe1,daoe2,jens,white1,white2}.
In a related line of work, Ref.~\cite{parker} focused on the the Liouvillians $\mathscr{L}$ of many-body systems. They examined the form of this operator with respect to its Krylov basis (the basis associated with a Lanczos procedure). In this basis the Liouvillian is tridiagonal. The off-diagonal elements are a sequence of positive numbers ${b_n}$ called the Lanczos coefficients. They formulated the operator growth hypothesis, which states that the Lanczos coefficients generically grow as $b_n\sim n$ asymptotically (up to logarithmic corrections in 1D). The rate of linear growth gives a bound on the growth of various notions of operator complexity (such as OTOCs), the growth of moments of form $\mu_n\equiv (\mathcal{O}|\mathscr{L}^n|\mathcal{O})$, and on the high frequency decay of local correlation functions $C(t)\equiv (\mathcal{O}|e^{i\mathscr{L}t}|\mathcal{O})$  which can be measured experimentally \cite{exp2,exp3}. Lastly, it appears that OGH can be used as part of an efficient numerical algorithm for approximating transport coefficients in high temperature quantum systems. 

While OGH is  connected to the growth of operator complexity with time, its connection to ergodicity is less clear. More recent works suggest that OGH can obtain in systems which may well be non-ergodic \cite{Cao,trigueros}. A possible scenario is that ergodicity implies OGH, but not the other way around. It could thus be that OGH holds for sufficiently generic, but possibly non-ergodic many body systems. It could also be that more refined versions of OGH discriminate between ergodic and non-ergodic systems \cite{Cao,trigueros}. Delineating the circumstances in which OGH and its variants obtain is useful as it will clarify the connection between operator complexity growth and ergodicity; it may also inspire improvements to the above-mentioned numerical algorithms for calculating transport coefficients, and enable us to apply these techniques to a wider class of models.

In this work we investigate the validity of OGH in \emph{open} quantum systems described by Lindbladian evolution \cite{Lindblad1976,book1,book2,lidar}. Modern quantum simulators are able to implement increasingly sophisticated dissipative channels \cite{midcircuitFeig,midcircuitIBM}, so it is worth understanding how OGH generalises to open quantum systems and to understand whether its generalisation corresponds to experimentally observable features. Moreover, even closed (ergodic) quantum systems can behave as baths for themselves, so we might expect features of open-system OGH to obtain in suitable variables in closed systems.

Some existing studies have addressed variants of this problem. In an attempt to generalize OGH to open systems, Ref.~\cite{liu} maps the Krylov complexity in an open system to a non-Hermitian tight binding model in similar spirit to Ref.~\cite{parker}. They consider the Lanczos basis defined for closed systems and conjecture that with Hermitian dissipation, the matrix elements of the dissipative part of the Lindbladian are dominated by the diagonal terms that grow linearly which they test in finite sized SYK and 1D interacting Fermionic models. Alternatively, Ref.~\cite{Bhat1} uses the Arnoldi approach that results in an upper Hessenberg (rather than tri-diagonal) form for the Lindbladian; their small system size numerics are consistent with a linear growth in these matrix elements. Likewise, Ref.~\cite{Bhat2} studies the Lanczos coefficients in the dissipative $q$-body SYK model also using the Arnoldi method. In the $q\to \infty$ limit, they gain analytic advantage and show that the Lindbladian is tri-diagonal with the Lanczos coefficients growing asymptotically linearly similar to closed system OGH. For finite $q$, they report linear growth in the first few diagonal and primary off-diagonal Lanczos coefficients of the upper Hessenberg matrix. Finally, Ref.~\cite{Bhat3} makes use of the so-called bi-orthonormal scheme (explained below) to numerically study the Lanczos coefficients in small systems, focusing on their behavior at very large $n$ compared to system size, and in doing so distinguish open chaotic and integrable behavior.\color{black}

In the present work, we focus on the bi-orthogonal scheme; we wish to understand the behavior of Lanczos coefficients at large $n$ but in the thermodynamic limit (where system size has been taken to infinity first, in contrast to \cite{Bhat3}) -- this is the regime in which the original OGH applies. Our results are as follows. We begin with the Lindbladian evolution for an operator $\hat{O}$ 
\begin{equation}\label{lind}
\frac{d\hat{O}}{dt}=\hat{\mathscr{L}}\hat{O}=i[\hat{H},\hat{O}]+\eta\sum_i[\hat{h}^{\dagger}_i\hat{O}\hat{h}_i-\frac{1}{2}\{\hat{h}^{\dagger}_i\hat{h}_i,\hat{O}\}],
\end{equation}
where $H$ is the Hamiltonian, $h_i$ are dephasing Hermitian jump operators with strength $\eta$ \cite{Lindblad1976,book1,book2,lidar}. In exchanging a Liouvillian for a Lindbladian, two important changes arise. It is useful to examine them from the perspective of the moments defined above $\widetilde{\mu_n}=(O|{\mathscr{L}}^n|O)$. Firstly, unlike closed systems, for open systems, the moments can be non-zero for odd values of $n$. This has an important consequence for the high frequency part of the spectral function which decays as a power-law in contrast to an exponential decay for closed systems in the original OGH. Secondly, dissipation affects the growth of moments Fig.~\ref{toy2}. While in unitary systems the moments tend to grow as $O((2n)!)$, in an open system there are strong deviations and oscillations in the value of the moments which kick in at $n\sim 1/\eta$. We demonstrate this result with numerical results on spin-chains and semi-analytic results in the dissipative SYK model. We are also able to reproduce the key features in the moments analytically by constructing a toy model for the Lindbladian as a non-Hermitian hopping model on operator space described in Fig.~\ref{toy1} which is somewhat similar to the construction in Ref.~\cite{liu}\color{black}. 

We derive a relation (Eq.~\ref{mutob}) between the moments and the Lanczos coefficients obtained through the bi-orthonormal scheme. We find that the point at which the moments start to deviate from the unitary scaling coincides with a dramatic change in the Lanczos coefficients: The Lanczos coefficients grow linearly initially until $n\sim 1/\eta$ at which point they experience exponentially growing oscillations. We are able to confirm this behavior in the large $q$ dissipative SYK model where we have a partial analytic control and numerically in the chaotic Ising model. Our toy non-Hermitian hopping model also reproduces this behavior in the Lanczos coefficients. 

Finally, due to the non-vanishing odd moments, the spectral function initially decays exponentially in frequency before crossing over to power law decay at a frequency set by the dissipation. We provide an explicit asymptotic formula for the crossover frequency in the case of the dissipative SYK chain Eq.~\ref{crsovr} where this frequency scales as $\sim \log(\eta^{-1})$, an asymptotic form we expect will hold more generally. This crossover and consequent power-law decay can be interpreted as a signature of modified OGH for an open Lindblad system, and sharply distinguishes open experimental systems subject to effective Lindbladian dynamics from closed systems. Moreover, systems in contact with \emph{equilibrium} baths can be viewed as larger closed systems with ergodic dynamics, therefore we expect their spectral functions to decay exponentially with frequency. Our results therefore suggest that local Lindbladians are never a good model for systems in contact with equilibrium baths, at least when one scrutinises the high-frequency behavior of spectral functions. 

\section{Bi-Lanczos approach}
Here we put the Lindbladian from Eq.~\ref{lind} in tri-diagonal form by constructing a so-called  bi-orthonormal Lanczos basis. In the absence of dissipation, this procedure coincides with the original Lanczos recipe in Ref.~\cite{parker}. We start by choosing our initial  Hermitian operator to be $|\mathcal{O}_0):=|\mathcal{O})$, normalized with respect to the \emph{infinite temperature inner product} $(X|Y):= \Tr[X^{\dagger}Y]/\Tr[I]$. The bi-orthonormal Lanczos basis is generated inductively \cite{GRUN}

\begin{equation}
\begin{split}
&|A_n): = (\mathscr{L}-a_{n-1})|\mathcal{O}_{n-1})-c_{n-1}|\mathcal{O}_{n-2})\\
&|B_n):=(\mathscr{L}^{\dagger}-a^{*}_{n-1})|\tilde{\mathcal{O}}_{n-1})-b_{n-1}|\tilde{\mathcal{O}}_{n-2}),
\end{split}\label{Krylov}
\end{equation}
where, 
\begin{equation}
\begin{split}
&|\mathcal{O}_{n}):=\frac{|A_n)}{b_n}\;\;\;\;\;(\tilde{\mathcal{O}}_{n}|:=\frac{(B_n|}{c_n}\\
& a_n:=(\tilde{\mathcal{O}}_n|\mathscr{L}|\mathcal{O}_n),\;\;\;b_n:=\sqrt{(A_n|A_n)},\;\;\;c_n:=\frac{(B_n|A_n)}{b_n}.\\
\end{split}
\end{equation}
The Lindbladian $\mathscr{L}$ is a tridiagonal matrix in this new basis.
Each iteration of the algorithm outputs three numbers $a_n,b_n,c_n$ and these make up the tridiagonal matrix elements

\begin{equation}
\mathscr{L}= \left( \begin{array}{cccc}
a_0 & b_1& & 0\\
c_1 & \ddots & \ddots & \\
& \ddots & \ddots & b_{n} \\
0 & & c_n& a_{n-1} \end{array} \right) .
\end{equation}
One can easily check that setting $\eta=0$ results in the usual Lanczos basis with $b_n=c_n$ and $a_n=0$. In this work we focus mainly on the growth of the sequence $\sqrt{b_n c_n}$ which we hereafter call the Lanczos coefficients.

\begin{figure}[t]
\includegraphics[width=\linewidth]{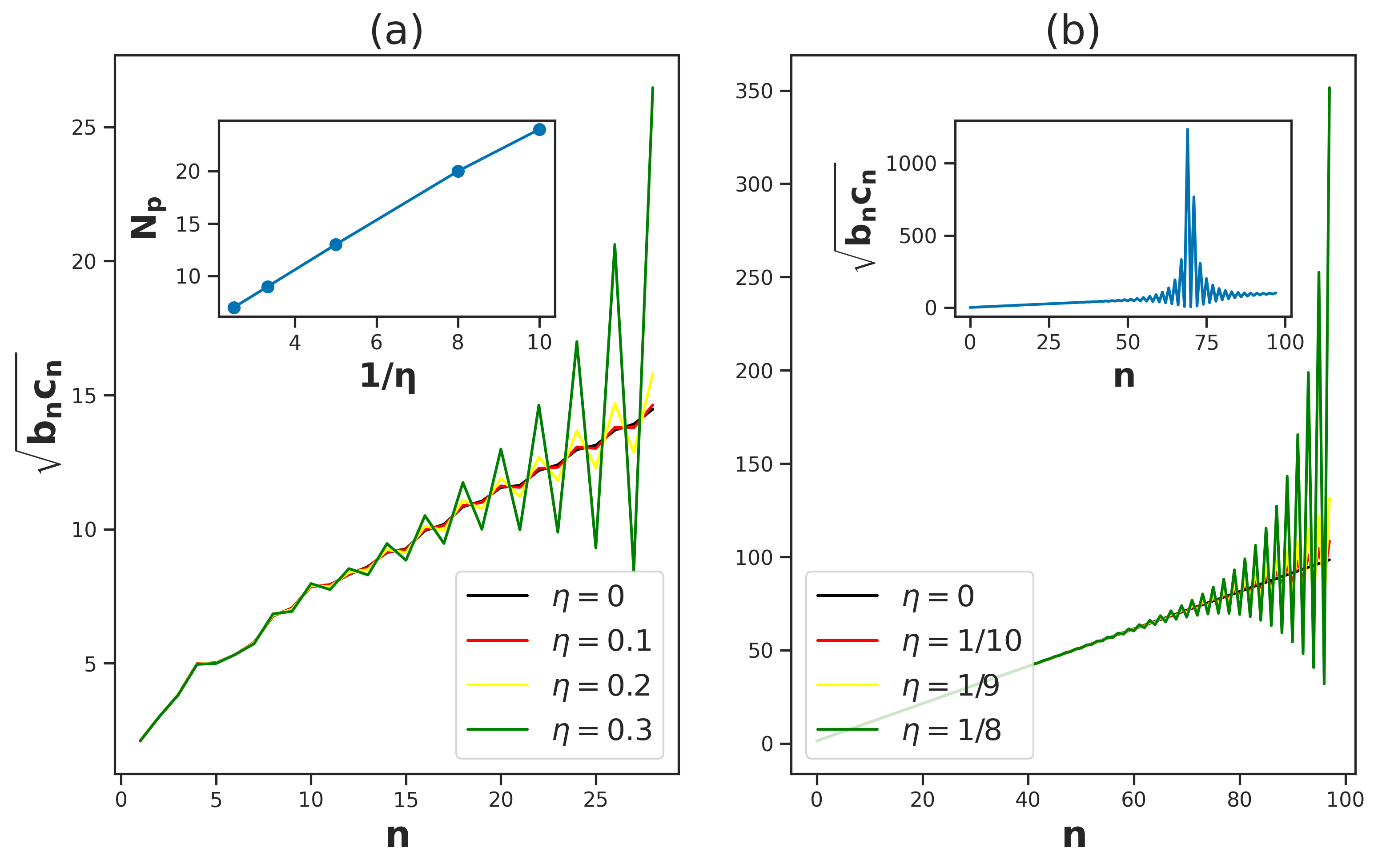}
\centering
\caption{Growth of Lanczos coefficients $\sqrt{b_nc_n}$ in (a) the dissipative Ising model and (b) the dissipative SYK model with $q=1000$ for different dissipation strengths $\eta$. We see an initial linear growth followed by oscillations whose amplitude grows exponentially with $n$. Inset in (a) shows that the point $N_p$ in the Lanczos iteration where $\sqrt{b_nc_n}$ begins to show deviation from linear behavior scales as $N_p\propto 1/\eta$ for the Ising model. The inset in (b) shows that for large $n$, Lanczos coefficients refocus to linear growth for the SYK model.}\label{Lan} 
\end{figure}

The Lanczos coefficients are also encoded, albeit indirectly, in the autocorrelation function
\begin{equation}\label{corfun}
C(t)\equiv(\mathcal{O}|e^{i\mathscr{L}t}|\mathcal{O}).
\end{equation}
$C(t)$ admits a Taylor expansion about $t=0$ in terms of the moments $\mu_n$ given by,
\begin{equation}
C(t)=\sum_{n=0}^{\infty}\mu_{n}\frac{(it)^n}{n!}.\label{cor}
\end{equation}
The Lanczos coefficients $b_n c_n$ are related to the moments through
\begin{equation}\label{mutob}
b_n c_n=\frac{K_n K_{n-2}}{(K_{n-1})^2},
\end{equation}
where $K_n=\det[\mu_{i+j}]_{0\leq i,j \leq n} $ and $K_{-1}=K_{0}=1$. The proof of this is provided in the App.~\ref{proof}.

We now study the growth of local operators in the bi-orthonormal Lanczos basis in systems whose unitary dynamics is known to be quantum chaotic. We begin by adding dissipation to an SYK model, then do the same for an ergodic Ising chain, before finally constructing a toy model which captures qualitatively the behaviour of all systems studied.

\subsection{Dissipative SYK model}
\begin{figure}[t]
\includegraphics[width=\linewidth]{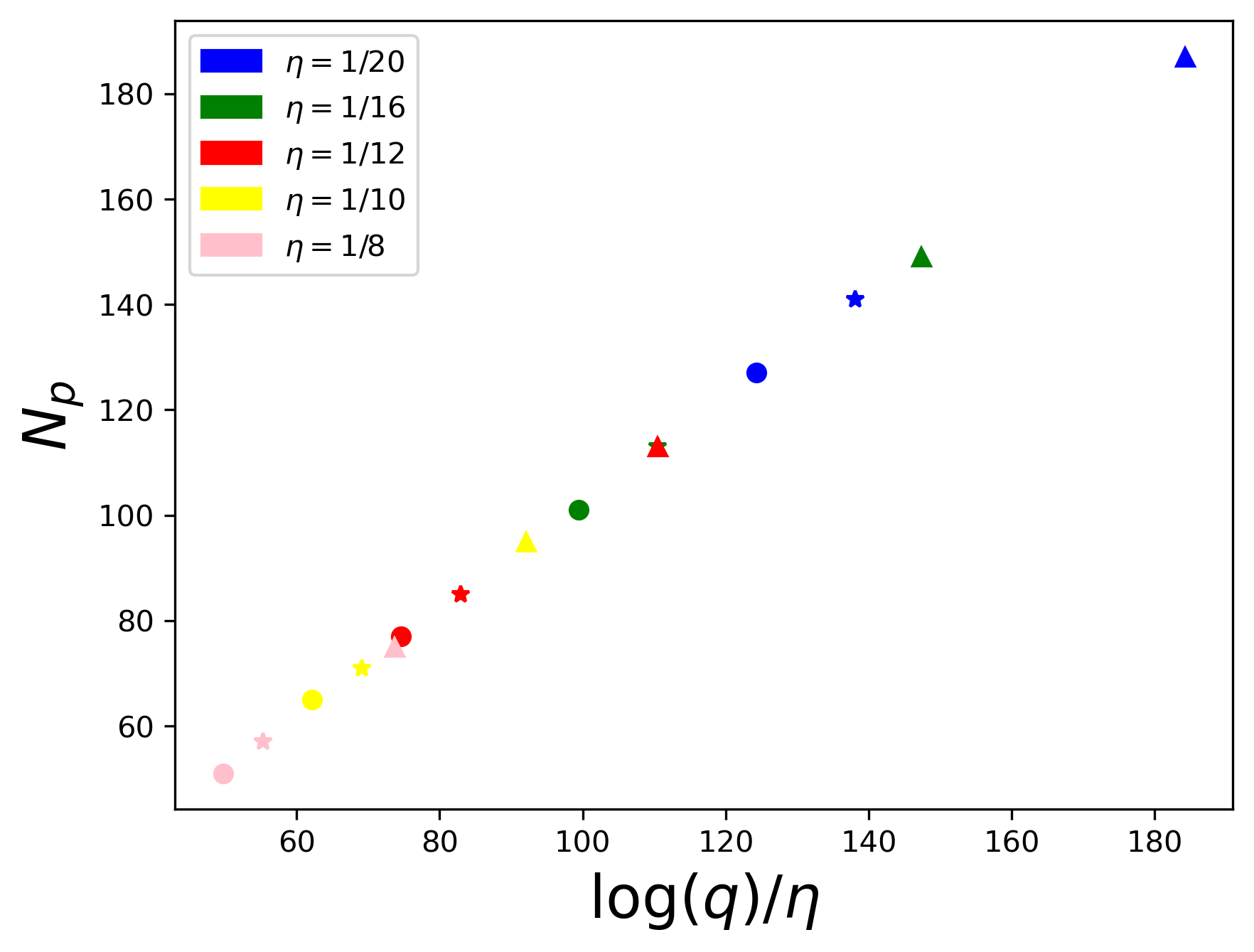}
\centering
\caption{The point $N_p$ where the Lanczos coefficients deviate from linear growth behavior for the dissipative SYK model computed by checking for the condition $|b_nc_n-n(n-1)|>\epsilon$ where we compute $b_nc_n$ numerically from Eq.~\ref{mutob}. We have chosen $\epsilon=1$ and have set $J=1$. We consider different choices of $q =5\times 10^2,10^3,10^4$ (marked by circle, star and triangle symbols respectively) and $\eta$ (with different colors) and plot $N_p$ against $\log(q)/\eta$. We see a linear scaling that agrees with the prediction from Eq.~\ref{lanc}}.\label{sykp}
\end{figure}

We start with the standard SYK model with $q$-body interactions given by \cite{Ye,Rosenhaus}
\begin{equation}
H_{SYK}=i^{q+1}\sum_{i_1<...<i_q} J_{i_1...i_q}\gamma_{i_1}...\gamma_{i_q}
\end{equation}
where $J_{i_1...i_q}$ are random couplings drawn from the Gaussian distribution and $\gamma_i$ are the Majorana operators. We add dissipative terms (with strength $\eta$) through the $N$ Lindblad jump operators that are linear in the Majoranas given by,
\begin{equation}
h_i=\sqrt{\eta}\gamma_i, \;\;i=1,...N
\end{equation}
Our starting point is the correlation function described in Eq.~\ref{corfun} for a single Majorana operator $|\mathcal{O})=\gamma_1$. This can be analytically computed in the large $q$ limit and is given by \cite{Anish,Tomaz},
\begin{equation}
C(t)=1+\frac{2}{q}\ln{\sech(\alpha t+\beta)},\;\;\;\;t>0 \label{corfun}
\end{equation}
where, $\alpha=J\sqrt{(\frac{\eta}{2 J})^2+1}$ and $\beta=\arcsinh (\frac{\eta}{2J})$. Notice that setting the dissipation $\eta=0$ results in the usual SYK correlation function \cite{Maldacena}. One can  compute the moments $\mu_n$ that appear in the Taylor expansion of $C(t)$ (Eq.~\ref{cor}) through the expression,
\begin{equation}
\mu_n=\frac{1}{i^n}\frac{d^n}{d t^n}C(t)|_{t=0}
\end{equation}
\\
Using this formula,  Eq.~\ref{mutob}, and the analytical result in Eq.~\ref{corfun}, we can compute the Lanczos coefficients. The results are shown in Fig.~\ref{Lan}. For finite $q$ we see that there is an initial linear growth of $\sqrt{b_nc_n}$ followed by a regime of exponentially growing oscillations. Also, interestingly, the Lanczos coefficients refocus to linear growth behavior for large $n$ as shown in the inset of Fig.~\ref{Lan}.  We compute numerically the point $N_p$ where the Lanczos coefficients deviate from linear growth behavior by checking for the condition $|b_nc_n-n(n-1)|>\epsilon$, where $\epsilon$ is a small cutoff and we have set $J=1$. We find that $N_p$  scales as $\log q/\eta$ as demonstrated in Fig.~\ref{sykp}. 

Alternatively, Lanczos coefficients may also be expressed up to $\mathcal{O}(1/q)$ as,

\begin{widetext}
\begin{equation}\label{lanc}
b_nc_n=\left\{ 
  \begin{array}{ c l }
    2 J^2/q & \quad  n=1 \\
    J^2\Bigl[n(n-1)+\frac{1+(-1)^{n+1}[n\cosh{([2n-2]\beta)}+(n-1)\cosh{(2 n \beta})]}{q}\Bigr]+\mathcal{O}(\frac{1}{q^2})                 & \quad n>1
  \end{array}
\right.\\\\
\end{equation} 
\end{widetext}

This formula is obtained by analytically calculating the first
few Lanczos coefficients using the procedure outlined above.
In doing so, the formulae contain recognisable sequences of
integers. Our analytical formula appears to agree exactly with all of our numerical checks. We infer the following from Eq.~\ref{lanc}. Firstly, in the $q\to\infty$ limit, $\sqrt{b_nc_n}$ grows linearly which is in agreement with the findings in Ref.~\cite{Bhat2}. Secondly, the point $n=N_p$ where the growth of $\sqrt{b_nc_n}$ begins to deviate from the linear growth occurs when $e^{2N_{p}\beta}\approx q$ and hence, for small $\eta$, we have $N_p\propto \log{q}/\eta$ which is in agreement with our numerics as shown in Fig.~\ref{sykp}. Lastly, it explains the initial linear growth of the Lanczos coefficients that we observe numerically in Fig.~\ref{Lan}.

\subsection{Dissipative chaotic Ising model}
It is necessary to understand the behavior of the Lanczos coefficients in a local interacting model with dissipation and how it compares with the findings from the SYK model. We choose the tilted field Ising model which is known to be chaotic and which appears to obey the OGH \cite{parker}. We consider the Hamiltonian at the `maximally chaotic' point given by,
\begin{equation}
H=\sum_i\sigma^{x}_i\sigma^{x}_{i+1}-1.05\sigma^z_i+0.5\sigma^x_i.\label{ising}
\end{equation}
We introduce dissipation through the jump operators $h_i=\sigma^z_i$ with strength $\eta$. Our choice for the initial operator is $|\mathcal{O}_o)=\sum_{i}\sigma^x_i$. We apply the Lanczos approach defined in Eq.~\ref{Krylov} and study numerically the growth of Lanczos coefficients for different dissipation strengths $\eta$. We work in the thermodynamic limit and are able to compute tens of Lanczos coefficients before hitting a memory ceiling. Fig.~\ref{Lan} shows a plot of Lanczos coefficients for the Ising model and we infer the following similarities to the dissipative SYK model. Firstly, we see that with dissipation, there is an initial transient Lanczos period where there is an approximate linear growth in $\sqrt{b_nc_n}$ followed by oscillations that increase rapidly. Secondly, the point $N_p$ in the Lanczos iteration where the coefficients deviate from linear growth behavior scales with dissipation as $N_p \propto 1/\eta$. This suggests some degree of universality in the behavior of Lanczos coefficients in chaotic many-body systems in the presence of Hermitian dissipation.

\begin{figure}[t]
\includegraphics[width=\linewidth]{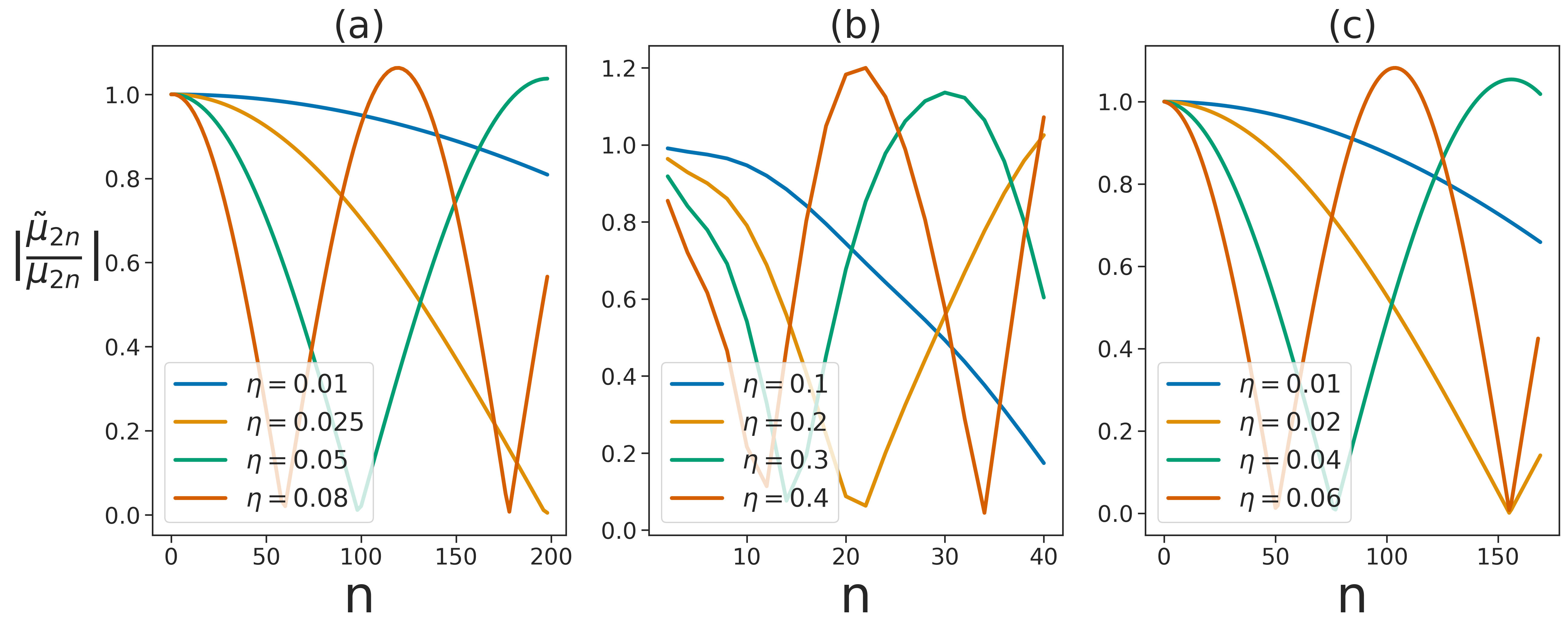}
\centering
\caption{Ratio $|\widetilde{\mu}_{2n}/\mu_{2n}|$ vs $n$ for different dissipation strengths $\eta$ plotted for a) The large $q$ SYK model with $q=500$ and $J=1$, b) The chaotic Ising model and c) The toy model with $J=1$. We see that dissipation impacts the growth of moments and the ratio $|\widetilde{\mu}_{2n}/\mu_{2n}|$ can get close to zero, indicating that $\widetilde{\mu}_{2n}$ flips sign.}\label{toy2}
\end{figure}

\section{Behavior of moments and the toy model}
Recall that the Lanczos coefficients are related to the moments through Eq.~\ref{mutob}, so that the breakdown of linear growth in the Lanczos coefficients should appear in the behavior of moments. To understand how dissipation affects the growth of moments, we study the ratio $|\frac{\widetilde{\mu}_{2n}}{\mu_{2n}}|$ (where $\widetilde{\mu}_{2n}$ and $\mu_{2n}$ are the moments with and without dissipation respectively) numerically for different dissipation strengths for the SYK and the dissipative Ising model as shown in Fig.~\ref{toy2}. We see that the ratio oscillates and in fact, for the Ising model, the point where the ratio first changes sign corresponds approximately to the point $N_p$ where the Lanczos coefficients show deviations from linear growth behavior seen in Fig.~\ref{Lan}. For the SYK model, the moments and the Lanczos coefficients depend on $q$ and $\eta$, but the ratio $|\frac{\widetilde{\mu}_{2n}}{\mu_{2n}}|$ depends only on $\eta$. Hence, we do not expect the point where the ratio changes sign to correspond to the point where the Lanczos coefficients deviate from linear growth for the SYK model for a given choice of $\eta$ and $q$.

The behavior of the moments and the Lanczos coefficients in models with Hermitian dissipation must be encoded more generally in how the unitary and dissipative contributions from the Lindbladian weigh relatively. Under repeated application of the unitary ($\mathscr{L}_{\textrm{u}}$) and dissipative ($\mathscr{L}_{\textrm{d}}$)  parts of the Lindbladian on an operator $\mathcal{O}_n=\mathscr{L}^{n}_{\textrm{u}}\mathcal{O}$, for some local initial operator $\mathcal{O}$, we expect that,

\begin{equation}\label{scaling}
||\mathscr{L}_{\textrm{u,d}}\mathcal{O}_n||^2\propto n^2||\mathcal{O}_n||^2
\end{equation}
The scaling for the unitary part comes from the growth of moments $\mu_{2n}=||\mathscr{L}^n_{\textrm{u}}\mathcal{O}||^2\approx \mathcal{O}((2n)!)$ for closed systems \cite{parker}. Interestingly, the dissipative part too has a similar scaling which may be understood by the following intuition. The operator $\mathcal{O}_n$ has a support on roughly $n$ sites and dephasing does not grow the operators which involve Pauli strings that are eigenstates of the jump operators \cite{schuster}. We hence expect every site to contribute under the action of $\mathscr{L}_{\textrm{d}}$ on the operator $\mathcal{O}_n$ which is the reason behind the scaling \cite{liu}. 

Although both the unitary and dissipative contributions have the similar scaling with $n$ when acting on the original Lanczos basis, they also involve contributions from the finite local bandwidth of the Hamiltonian and the dissipation strength respectively. Note too that the dissipative part of Eq.~\ref{lind} comes with a relative $i$. This suggests the following toy model for the action of the Lindbladian on the Lanczos basis for $\mathscr{L}_u$ defined as $|n\rangle:=\mathscr{L}^{n}_{\textrm{u}}\mathcal{O}$,

\begin{equation}\label{toym}
\begin{split}
&\mathscr{L}_{\textrm{u}}|n\rangle = J|n+1\rangle\\
&\mathscr{L}_{\textrm{d}}|n\rangle = (i\,\eta\, n)|n\rangle\\
\end{split}
\end{equation}
We approximate the moments for the unitary part with their anticipated asymptotic form $\mu_{2n}=\langle n|n\rangle=(2n)!$ \cite{parker}.  Our toy model allows us to express the modified moments $\widetilde{\mu}_n=\langle 0|(\mathscr{L_{\textrm{u}}}+\mathscr{L_{\textrm{d}}})^n|0\rangle$ in terms of the unitary moments $\mu_{2k}, k=1,2,..,\lfloor n/2 \rfloor$ as,

\begin{equation}\label{toymom}
\widetilde{\mu_{n}}=\sum^{\lfloor n/2 \rfloor}_{k=1}\stirling{n}{2k}\;(i\eta)^{n-2k}\;(2k)!
\end{equation}
where, $\stirling{n}{2 k}$ represent the Stirling numbers of the second kind and we have used $\langle 0|2k \rangle=\mu_{2k}$.

\begin{figure}[t]
\includegraphics[width=\linewidth]{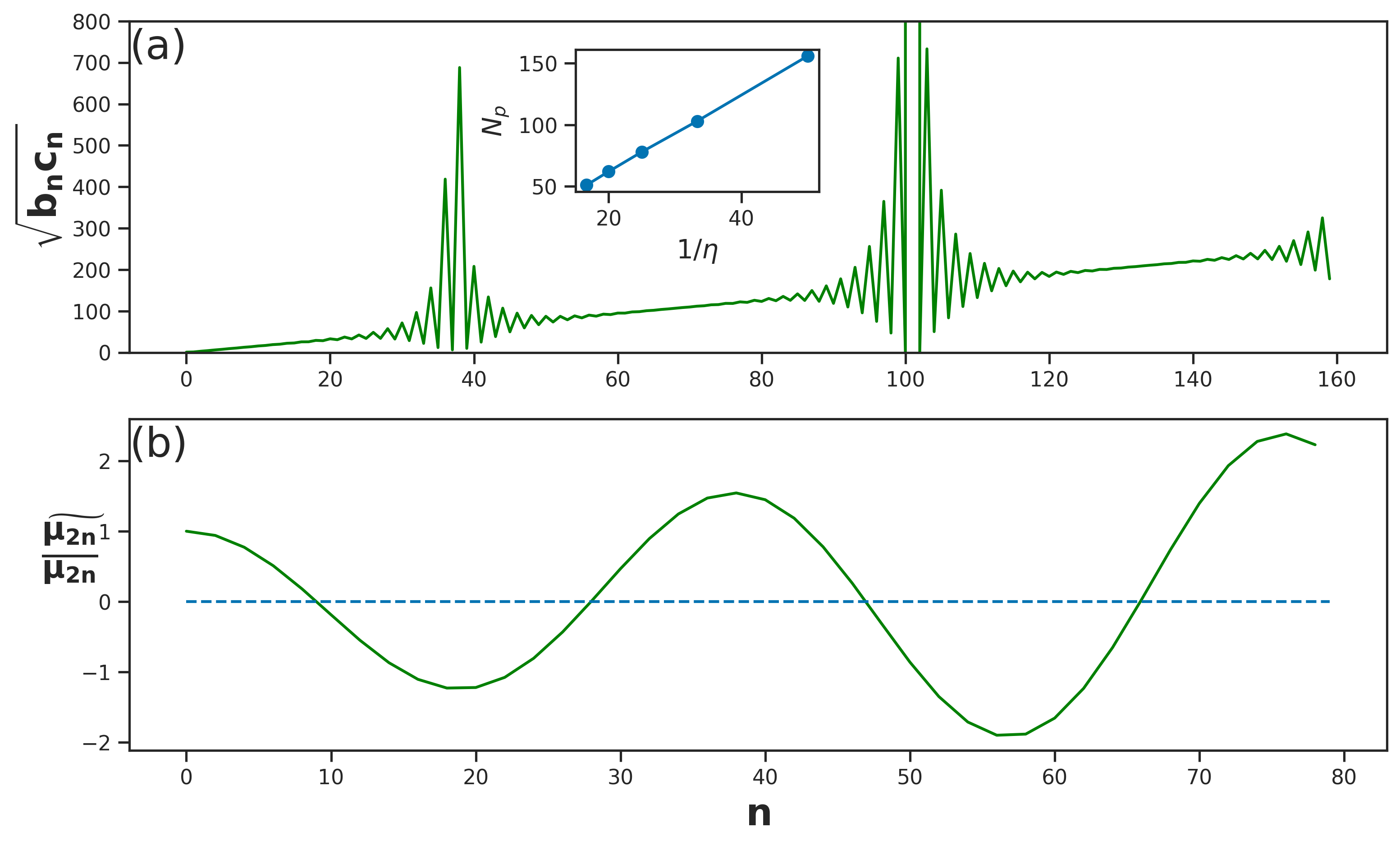}
\centering
\caption{Numerical results for the toy model. (a) Growth of Lanczos coefficients $\sqrt{b_nc_n}$. The Lanczos coefficients have an overall linear growth but with periods of oscillatory bursts. The inset shows that the point $N_p$ where $\sqrt{b_nc_n}$ first deviates from linear behavior scales as $N_p \propto 1/\eta$. (b) Ratio $\widetilde{\mu}_{2n}/\mu_{2n}$ of the even moments with and without dissipation plotted against $n$. It is seen that the Lanczos coefficients first deviate from linear growth at around $n\approx20$ which corresponds to the ratio between the moments crossing zero at around the same point. The oscillations in the moments explain the periodic behavior of oscillatory bursts in the Lanczos coefficients. For the main plots we have chosen $J=1$ and $\eta=1/6$. }\label{toy3}
\end{figure}

Remarkably, the toy model captures a similar oscillatory behavior for the ratio ($|\frac{\widetilde{\mu}_{2n}}{\mu_{2n}}|$) as shown in Fig.~\ref{toy2}. In the limit of small $n$ and $\eta$, we can approximate the ratio by the last two terms of the summation in Eq.~\ref{toymom} leading to,

\begin{equation}
|\frac{\widetilde{\mu}_{2n}}{\mu_{2n}}|\approx 1 - O(\eta^2 n^2)
\end{equation}
This suggests that the ratio gets close to zero when $n\sim 1/\eta$, which is in agreement with our results for the SYK and the Ising model. To understand in more detail why the  $\widetilde{\mu}_{n}$ fluctuates in sign, we examine our toy model expression Eq.~\ref{toymom}. The expression involves a sum of terms with alternating sign. But the sum tends to be narrowly peaked around a particular $k$ set by $n,\eta$. As $n$ is changed, the sign of the dominant term flips with a period that goes like $1/\eta$.

Next, we compute the Lanczos coefficients for the toy model from the moments $\widetilde{\mu}_n$ described in Eq.~\ref{toy3} and we see the expected initial behavior for $\sqrt{b_nc_n}$ : linear growth followed by growing oscillations as shown in Fig.~\ref{toy3} with the scaling $N_p \propto 1/\eta$ similar to the SYK and the Ising models. Interestingly, the toy model also shows that the Lanczos coefficients have an overall linear growth but suffers bursts of oscillations that grow and decay occurring at periods controlled by $1/\eta$. Finally, we also note that in the limit of very large dissipation, the Lanczos coefficients remain bounded and do not grow until the contribution from the unitary part becomes comparable to the dissipative part and we see this in both the dissipative Ising model and the toy model. This further validates that the effective description for the Lindbladian used in the toy model serves as a good approximation to the many-body problem.

\section{Spectral function and experiments} 
Following the OGH for closed systems, we turn our attention to the spectral function $\phi(\omega)$ which is an experimentally accessible quantity. We may define this as the cosine Fourier transform of $C(t)$ \footnote{It is preferable to use the cosine representation of the spectral function, as Lindbladian evolution is only guaranteed to be completely positive for positive times}

\begin{equation}
\phi(\omega)=F[C(t)]=\int^{\infty}_{0}dt\cos{(\omega t)}\,C(t).
\end{equation}
As mentioned earlier, for an open Lindblad system, in addition to even moments, odd moments are non-vanishing. This has an important consequence for the high-frequency behavior of the spectral function. For an auto-correlation function $C(t)$ that decays and possesses non-zero odd moments (i.e, not an even function of $t$), one can show that $\phi(\omega)$ decays as a power law. As an example, let us consider the resummed dissipative SYK correlation function \cite{stanford} given by
\begin{equation}\label{resummed}
C_{\textrm{R}}(t)=(\cosh{(\alpha t+\beta)})^{-2/q}.
\end{equation}
for $t\geq 0$. Here $\alpha$ and $\beta$ have the same definitions as in Eq.~\ref{corfun} and we set $J=1$ for simplicity.

\begin{figure}[t]
\includegraphics[width=\linewidth]{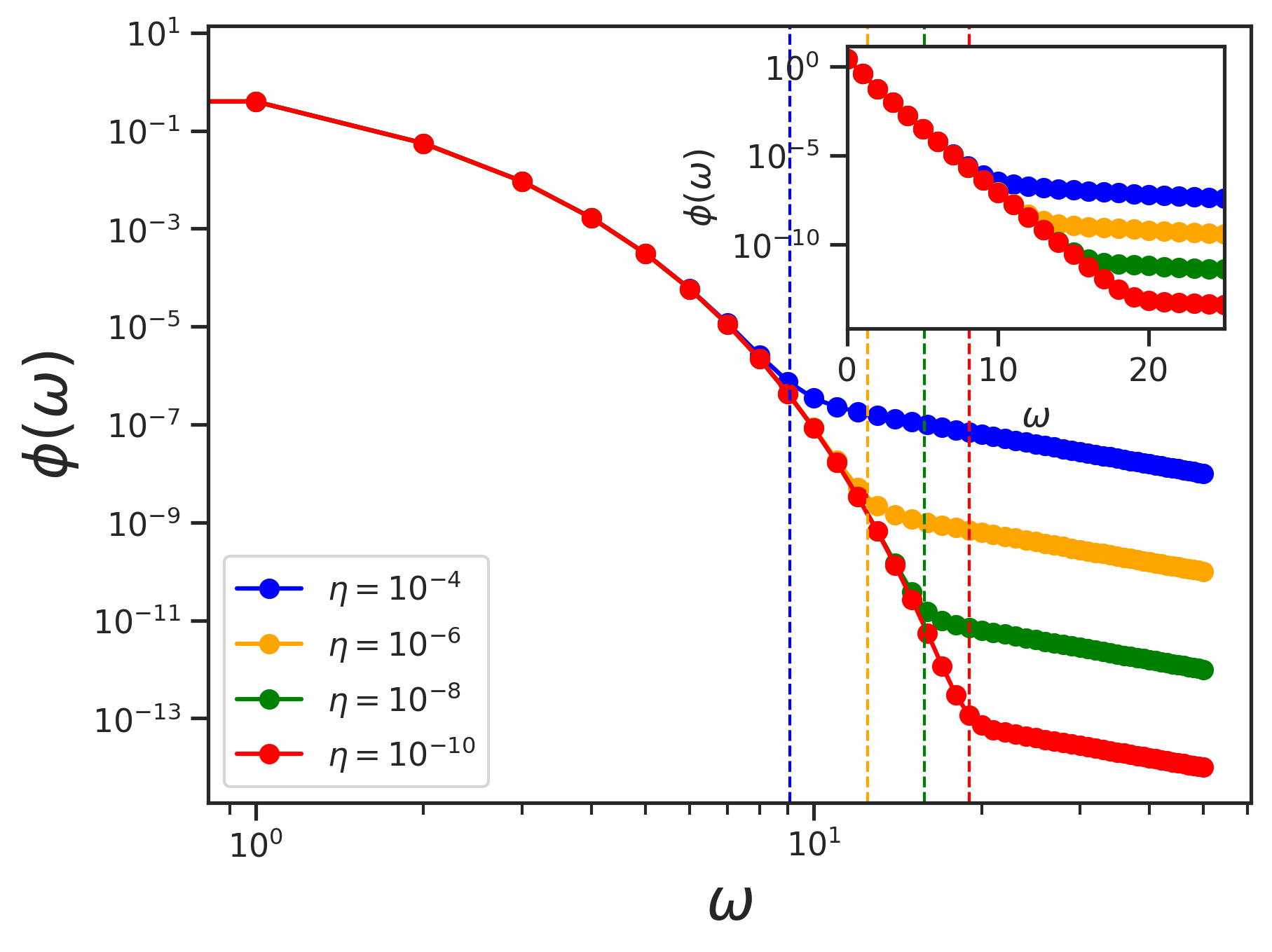}
\centering
\caption{Spectral function $\phi(\omega)$ computed numerically for the resummed dissipative SYK correlation described in Eq.~\ref{resummed} with $q=4$, $J=1$ and for different choices of dissipation strength $\eta$. A cross over from an exponential to a power-law decay is seen to occur roughly at the point $\omega^{*}$ computed from Eq.~\ref{crsovr} for each $\eta$ and is marked by  vertical dashed lines. The main plot is shown with a log-log scale and the plot in the inset is shown with a log-linear scale.}\label{specfn}
\end{figure}

We have the following relation for the cosine transform of the second derivative of the correlation function $C(t)$ given by
\begin{equation}\label{asym}
F[C^{''}(t)]=-\omega^2 F[C(t)]-\mathrm{i}\mu_1.
\end{equation}
Provided $C^{''}(t)$ and its derivatives decay fast enough in time, a Riemann-Lebesque style argument indicates that the LHS of Eq.~\ref{asym} decays at least as fast as $1/\omega$. This is indeed the case when $C=C_R$ ( provided $\alpha>0,\beta\geq 0$) but we expect it to hold more generally. In that case, at large $\omega$, the RHS of Eq.~\ref{asym}  tends to zero so that
\begin{equation}
\phi(\omega)=F[C(t)]\sim -\mathrm{i}\mu_1\omega^{-2}.
\end{equation}

Note that the power law tail arises if the first odd moment is non-vanishing $\mu_1\neq0$, and we only have non-vanishing odd moments when there is dissipation.  For example, in the  SYK case $C=C_R$, we see that the spectral function decays as $1/\omega^2$ at large $\omega$ precisely when $\beta\neq 0$. On the other hand, for the case without dissipation ($\beta=0$), we know that the SYK spectral function decays exponentially as $\phi(\omega)=e^{-\frac{\omega \pi}{2}}$ \cite{parker}. This suggests that the spectral function exhibits a crossover from an exponential to a power-law decay for an open system at high frequency. This is shown in Fig.~\ref{specfn} for the SYK model. For small dissipation $\eta$, we find approximately the following dependence on $\eta$ for the cross over frequency $\omega^*$ for the SYK model,
\begin{equation}\label{crsovr}
\omega^{*}\approx \frac{2}{\pi}\ln[\frac{4 \,q\, \eta^{-1}}{\pi^2}(\ln[\frac{4\, q\, \eta^{-1}}{\pi^2}])^2]
\end{equation}
We obtain the above expression by equating the expected power law decay ($-i\mu_1\omega^{-2}$) with the known exponentially decay ($e^{-\frac{\omega\pi}{2}}$) and solving for $\omega$. We find that the above formula is a good approximation to the actual cross-over frequency as shown in Fig.~\ref{specfn}.  We conjecture that the crossover between exponential and power-law decay should also obtain in more general ergodic systems obeying OGH to which weak dephasing has been added, and that the crossover frequency has approximate scaling $\omega^{*} \sim \ln(\eta^{-1})$ for the same reasons that lead to Eq.~\ref{crsovr}.

Spectral functions are readily measured in a variety of solid-state experiments. In NMR experiments these show an exponential decay with frequency agreeing with theoretical predictions for closed systems \cite{parker,exp1,exp4,exp5}. These experiments can also probe spin autocorrelation functions $C(t)$ by measuring the free induction decay (FID) signals from the nuclear spins. The first few even moments $\mu_{2n}$ can be extracted from such data \cite{exp2,exp3}, and there is no evidence that odd moments appear in these systems. Therefore, although the nuclear spins are part of an open system  coupled to a bath of electron spins and phonons, the spectral function data suggests they are in fact well-isolated from this bath. For example, in the FID experiments in $\textrm{CaF}_2$, the spin-lattice relaxation time is of the order of $10\mathrm{s}$ while the time scales associated with nuclear spin-spin interaction is of the order of $10^{-6}\mathrm{s}$ \cite{exp2}. This separation of scales occurs because the spin-lattice relaxation is mediated via interactions with electron spin degrees of freedom; this interaction is negligible in the non-magnetic materials studied \cite{slichter,abragam}. Hence, the unitary dynamics due to nuclear spin-spin interactions dominates the open-system dynamics due to nuclear-electron spin coupling and consequent spin-lattice relaxation. 

Modern quantum simulator experiments are a promising platform for testing open system OGH \cite{DiCandia,Del,Garcia,Schlimgen}. One can simulate quantum channels by embedding the non-unitary gate in a larger unitary gate acting on a system qubit and ancilla. In fact, the z-dephasing noise considered for the open Ising model above could be implemented even more simply by simulating a Hamiltonian system with a white noise and spatially uncorrelated  magnetic field on each site. This can be achieved by interspersing the Hamiltonian dynamics with random (in space and time) on-site z-rotations, and averaging the final results over a number of experimental runs. The spectral functions may then be measured by performing real-time linear response measurements, and cosine transforming the results to frequency space.

Lastly, we return to the idea that closed ergodic systems can behave as their own baths. For such setups, the reduced density matrices of subsystems tend towards a thermal state. It is prima facie plausible that these subsystem dynamics are themselves well-modelled by Lindbladians which drive them towards thermal equilibrium.  However, our results above suggest where the Lindbladian approximation breaks down. Consider a system made of two disjoint but extensive spatial components  $A,B$ (these could, e.g., be the sites on two disjoint sublattices of a parent lattice). For the full closed system dynamics, an observable based on $A$ presumably obey OGH and experience exponentially fast decay of their spectral functions. However, if the dynamics on $A$ is well-modelled by the OGH our results here (though limited to a subset of all Lindbladians), suggest  that this observable will obey power-law decay at high frequencies. These two predictions contradict one another, and show that the breakdown of the Lindbladian approximation is especially apparent in the high-frequency behavior of spectral functions. However, this disparity vanishes if  one accepts Lindbladians only give an accurate description of an equilibrating many body system when it is coarse-grained over time.

\section{Conclusions}
We have shown that open systems described by a Lindbladian with dephasing obey a modified version of the operator growth hypothesis. The Lanczos coefficients resulting from a bi-orthonormal Lanczos procedure for a Lindbladian initially grow linearly as in the closed system OGH but suffers exponentially growing oscillations whose occurrence is controlled by the inverse dissipation strength. We demonstrated this in the partially analytically solvable large $q$ SYK model and in the chaotic Ising model with dissipation introduced through dephasing jump operators. This behaviour is intimately connected to competing contributions from the unitary and dissipative components under repeated application of the Lindbladian. This motivated us to construct a simplified non-Hermitian toy model that successfully captures the behaviour of operator growth moments and the Lanczos coefficients together with the correct scaling with dissipation strength observed in the models studied. Finally, we showed that modified OGH for an open system manifests as high frequency power-law tails in the spectral function. This modified decay is a sharp experimental diagnostic distinguishing open system dynamics from closed-system unitary dynamics.

Our work hints at several worthy open questions. Firstly, we have restricted our attention to Lindbladians which have Hermitian jump operators, which always have an infinite temperature steady-state. It is an open question whether the results here generalise to a wider class of Lindbladians, namely those with non-Hermitian jump operators and more exotic steady states.  Next, the OGH is used in \cite{parker} to efficiently calculate transport coefficients. The instability of the Lanczos coefficients observed in this work suggests that a similar approach will \emph{not} work for open systems, at least using the bi-orthogonal  Lanczos scheme. It would be good to confirm this intuition, and to test whether the method similarly breaks down in alternative (e.g., Arnoldi) schemes.

\begin{acknowledgments}
The authors thank Ehud Altman, Oliver Lunt, Mingee Chung and Grigorii Starkov for insightful discussions. Both authors are supported by a UKRI Future Leaders Fellowship MR/T040947/1.
\end{acknowledgments}

\appendix
\section{Moments to Lanczos coefficients}\label{proof}
 
We establish how one can extract the Lanczos coefficients $b_nc_n$ starting from the moments $\mu_n$. If $|\mathcal{O}_0)$ is our initial normalized operator, then, $\mu_{n}=(\mathcal{O}_0|L^n|\mathcal{O}_0)$. We define,
\begin{equation}
\begin{split}
&|T_j):=L^{j}|\mathcal{O}_0)\;\;\;
(\tilde{T}_j|:=(\mathcal{O}_{0}|L^j\\
&M_{ij}:=(\tilde{T}_i|T_j)=(\mathcal{O}_o|L^{i+j}|\mathcal{O}_0)=\mu_{i+j}.\\
\end{split}
\end{equation}
The Krylov basis ($|\mathcal{O}_n), (\tilde{\mathcal{O}}_n|$) defined in \eqref{Krylov} is related to [$|T_j),(\tilde{T}_j|$] by
\begin{equation}
\begin{split}
&|\mathcal{O}_j)=R_{jk}|T_k)\\&(\tilde{\mathcal{O}}_j|=\tilde{R}^{*}_{jk}(\tilde{T}_k|.\\
\end{split}\label{trans}
\end{equation}
where the repeated indices are summed over.
Comparing \eqref{trans} and \eqref{Krylov} term by term and using the fact that $\det M=\frac{1}{\det \tilde{R}^{*}\det R}$, we find

\begin{equation}
b_nc_n=\frac{K_n K_{n-2}}{(K_{n-1})^2}
\end{equation}
where, $K_n=\det[\mu_{i+j}]_{0\leq i,j \leq n} $ and $K_{-1}=K_{0}=1$.

\end{document}